\documentclass[12pt,draftclsnofoot,onecolumn,doublespaced]{IEEEtran}
\ifCLASSINFOpdf
\else
\fi
\usepackage{cite}
\usepackage{graphicx}
\usepackage{amsmath}
\usepackage{amsfonts}
\usepackage{array}
\usepackage[tight]{subfigure}
\usepackage{amssymb}
\usepackage{amscd}
\usepackage{latexsym}
\usepackage{multirow}
\usepackage{graphics}
\usepackage{psfig}
\usepackage{epsfig}
\usepackage{color}
\usepackage{cite,algorithmic,algorithm}
\usepackage{amscd}
\usepackage{makeidx}
\usepackage{url}

\usepackage{multicol}
\usepackage{longtable}
\usepackage{lipsum}% just to automatically generate some text
\usepackage{supertabular}

\date{}

\newcommand{\be}{\begin{equation}}
\newcommand{\ba}{\begin{array}{lcl}}
\newcommand{\ea}{\end{array}}
\newcommand{\ee}{\end{equation}}
\newcommand{\ben}{\begin{equation*}}
\newcommand{\een}{\end{equation*}}
\newcommand{\bsp}{\mbox{ }}
\newcommand{\noi}{\noindent}

\newcommand{\bs}{\begin{small}}
\newcommand{\es}{\end{small}}

\newtheorem{lemma}{Lemma}

\newtheorem{theorem}{Theorem}

\title{
\vspace{-3cm}
\hfill {\em{\small Submitted to IEEE Trans. on Signal Processing}}\\
Exploiting Sparse Dynamics For Bandwidth Reduction In Cooperative Sensing Systems\thanks{*A preliminary version with a subset of the results was presented at the {\em Allerton Conference on Communication, Control and Computing}, Monticello, IL, Sep. 2010.}}

\author{Harish~Ganapathy, Constantine Caramanis and Lei Ying\thanks{H. Ganapathy and C. Caramanis are with the Department of Electrical and Computer Engineering, The University of Texas, Austin, TX 78712. L. Ying is with the School of Electrical, Computer and Energy Engineering, Arizona State University, Tempe, AZ 85286, USA.}\\
E-mail: \texttt{harishg@utexas.edu,caramanis@mail.utexas.edu,\\ lei.ying.2@asu.edu}}

\begin{document}

\maketitle
\begin{abstract}
Recently, there has been a significant interest in developing \emph{cooperative sensing} systems for certain types of wireless applications. In such systems, a group of sensing nodes periodically collect measurements about the signals being observed in the given geographical region and transmit these measurements to a central node, which in turn processes this information to recover the signals. For example, in cognitive radio networks, the signals of interest are those generated by the primary transmitters and the sensing nodes are the secondary users. In such networks, it is critically important to be able to reliably determine the presence or absence of primary transmitters in order to avoid causing interference. The standard approach to transmit these measurements from sensor the nodes to the fusion center has been to use orthogonal channels. Such an approach quickly places a burden on the control-channel-capacity of the network that would scale linearly in the number of cooperating sensing nodes. In this paper, we show that as long as one condition is satisfied: the {\it dynamics of the observed signals are sparse}, i.e., the observed signals do not change their values very rapidly in relation to the time-scale at which the measurements are collected, we can significantly reduce the control bandwidth of the system while achieving the full (linear) bandwidth performance.
\end{abstract}

\thispagestyle{empty}

\section{Introduction}\label{sec: intro}
Cooperative sensing is a promising technique that has received a lot of attention recently due to necessity for reliable decision-making. Here, a group of wireless nodes collects measurements about the some signals of interest. The observations are then reported to a central node, which in turn applies some appropriately-chosen decision rule to recover the signals. Such an architecture finds application in two closely-related and well-studied settings.

The first setting is (de-)centralized detection in sensor networks -- a group of sensor nodes commanded by a fusion center -- where the observed signals could be temperature or humidity readings, military targets, etc. Much work (see \cite{varshney,tomluo,spanias,veeravalli} and references therein), both theoretical and practical, has studied and optimized various aspects of the cooperative sensing architecture, e.g., quantization functions at the sensing nodes, the combining process and decision metric at the fusion center, the communication scheme between the sensing nodes and the fusion center, etc. This is a rich area of research and hence, our references are far from a comprehensive list. The second setting is a cognitive radio network where the observed signals are other unlicensed (secondary) or licensed (primary) transmitters. While the ideas in this paper are applicable to both the aforementioned settings, we review some literature from the latter as it is most relevant to the technical developments proposed here.

\subsection{Spectrum sensing in cognitive radio networks}
Cognitive radio technology, introduced by J. Mitola \cite{mitola}, is a promising solution to the growing scarcity of wireless spectrum, one that can potentially increase the spectrum utilization efficiency as recognized by the FCC \cite{FCC1,FCC2}.  Traditionally, a portion of spectrum is allocated or licensed for use by a specific group of users by regulatory agencies. On the other hand, cognitive radio networks call for cognitive (unlicensed/secondary) users to operate on the same frequency band as the primary licensed users while attempting to access the spectrum {\em seamlessly}. In other words, the cognitive users need to adjust their operating parameters to guarantee minimal impact to the primary licensed users. For example, DARPA's \emph{Next Generation} program \cite{darpa} has been interested in developing spectrum sensing techniques that prevent interference to existing occupants of the frequency band. To increase spectral efficiency, the FCC recently opened up TV whitespaces ($54$ MHz - $806$ MHz) for unlicensed use \cite{FCCruling}. While the ruling calls for access to a central database to determine TV band availability, the traditional use-case for spectrum sensing, we believe it still plays an important role in providing acceptable quality-of-service (QoS) over unlicensed bands. For example, a database provider could provide multiple classes of service to a \emph{querying} secondary user where the higher class would be provided with a per-TV-band estimate of the interference statistics from other unlicensed users in the band. Spectrum sensing might be necessary in this context in order to collect such statistics.

Cooperative or collaborative sensing in the context of cognitive radio networks consists of a group of secondary users or specially-placed sensor nodes that collect measurements over some sensing time window about the activity of the primary user and transmit these measurements to a central node. The central node or fusion center may often be the cognitive base station. Cooperative sensing techniques can be classified appropriately based on the type of decision metric used at the fusion center. We focus on one of the simplest schemes that relies on energy detection. Most prior work on energy-detection-based cooperative sensing (see \cite{ganesan,peh,sayeed,letaief,lee} and references therein), the signal is typically a single binary hypothesis modeling a system with one primary user that is either ON or OFF. In cooperative sensing, the measurements from each cognitive sensor node are typically linearly weighted and combined to form a decision statistic that is in turn compared against a threshold to produce a binary output. The paper by Ganesan and Li \cite{ganesan} constitutes one of the earliest contributions that establish the gains in detection (of a single primary user) probability due to cooperation in the presence of channel fading. Peh et al. \cite{peh} and Lee \cite{lee} study the tradeoff between sensing time and throughput of the cognitive network since longer sensing times lead to higher primary detection probabilities but lower secondary throughputs. Quan et al. \cite{sayeed} further optimize the linear weighted combiner studied by Ganesan and Li \cite{ganesan} by choosing weights that maximize the detection probability. While the above authors designed \emph{soft-combining} systems where the sensors typically report their measurements to the fusion center without any processing, Li et al. \cite{letaief} consider \emph{hard-combining} where local decisions are made at each sensor, and design a voting rule to produce the final binary decision. Finally, Mishra et al. \cite{sahai} compare the performance of hard-/soft-combining and study the effects of correlated fading on the detection probability. We refer the reader to a recent survey papers by Akyildiz et al. \cite{akyildiz} and Yucek \cite{yucek} for a more comprehensive list of references along with further discussion on the other types of cooperative spectrum sensing such as cyclostationary feature detection.

\subsection{Our contributions}
In almost all of the reviewed literature, the standard approach to transmit the measurements from the sensor nodes to the fusion center has been to employ orthogonal channels. This means that the control bandwidth demanded by the standard cooperative sensing scheme scales linearly in the number of sensing nodes. This immediately places a significant burden on the bandwidth requirements of the network and might be impractical in some scenarios as recognized by Akyildiz et al. \cite{akyildiz}. In this work, we focus reducing the amount of control bandwidth required by the system by considering the structure that may often exist in the \emph{dynamics} of the observed signals. Thus, we immediately build on and extend past models by adding a temporal dimension to the observed signal and by considering multiple discrete signals instead of a single binary hypothesis.

The key idea we exploit in this paper is the following: while the observed signals at a snapshot in time in general lives in some arbitrary high dimension, often, there is a time-scale separation between the sensing time window and the behavior of the observed process. For example, TV transmitters would turn ON/OFF on a significantly slower time-scale (in the order of minutes) than sensor measurement windows (in the order of milliseconds). Under these circumstances, {\em the dynamics of the observed signal vector is likely to be sparse}. It has long been known, and recently popularized under the name of compressive sampling, that whereas $N$ linear measurements are required to reconstruct a vector or signal in $\mathbb{R}^N$, if it is $S$-sparse (i.e., it has $S$ non-zero coefficients) then under appropriate conditions on the measurements, $\mathcal{O}(S \log N)$ are enough \cite{candesSigProcMag,baraniuk,donoho1}. By developing similar tools, and applying them on the dynamics, rather than the signal directly, we show that with greatly-reduced control bandwidth, our algorithms perform close to the linear control bandwidth case. The performance is measured in terms of the distance between the recovered signal and the true signal.

To the best of our knowledge, this is the first work in the space of wireless networks\footnote{Sparse changes in the dynamics of the signal have been used in other areas such as image processing to achieve video compression \cite{Richardson}.} to exploit {\em sparsity in the dynamics} of the observed signal. As this is likely much more prevalent than sparsity in the actual trajectory of the signal (of course, if the trajectory is sparse, then so are the dynamics) we expect this high-level idea to find broad application. More concretely, the main contributions in this paper are as follows:

\begin{enumerate}
\item[(1)] A first (to the best of our knowledge) application of compressive sampling to reduce the control bandwidth in a cooperative sensing system.

\item[(2)] A proof that path-loss matrices satisfy the \textit{null space property} thereby allowing for efficient acquisition or sensing of the observed signal using standard convex programs such as $\ell_1$-norm minimization and Lasso \cite{lasso}. The proof technique is novel since path-loss matrices contain entries that have non-zero mean and are not independent, a scenario that has not been dealt with extensively in past research.

\item[(3)] Simulation results that establish the competitive performance of our algorithm in comparison to the full control bandwidth case.
\end{enumerate}

The rest of this paper is organized as follows. In Section~\ref{sec: sysmodel}, we introduce  the system model for the cooperative sensing network under consideration. In Section \ref{sec: CS}, we discuss the compressive sampling algorithm that enables cooperative sensing using significantly-reduced control overhead. We establish the ``goodness'' of path-loss sensing matrices in Section \ref{sec: NSPPL}. The complete cooperative sensing algorithm is presented in Section \ref{sec: coopsensing}. Simulation results establishing the competitive performance of the algorithm are contained in Section \ref{sec: simu}. Concluding remarks are made in Section~\ref{sec: conclusion}.\\

\noi {\em Notation}: $x_{ij}$ denotes element $(i,j)$ of matrix ${\bf X}$ while $x_{i}$ denotes element $i$ of vector ${\bf x}$. $(.)^T$ is the transpose operator. For ${\bf x}\in{\mathbb R}^{N}$, ${\bf x}_{{\mathcal A}}$, ${\mathcal A}\subseteq \{1,2,\ldots,N\}$ denotes the vector ${\bf x}$ restricted to the entries in ${\mathcal A}$. For ${\bf X}\in{\mathbb R}^{m\times n}$, ${\bf X}_{{\mathcal A}}$, ${\mathcal A}\subseteq \{1,2,\ldots,m\}$ denotes the sub-matrix of ${\bf X}$ formed by the rows contained in ${\mathcal A}$. Finally, $||\cdot||_p$ is the $p$-norm operator on vectors.

\section{System model}\label{sec: sysmodel}
We consider a network with $N_s$ sensing nodes and a single fusion center operating in slotted-time. We introduce the necessary measurement and signal models below.

{\bf Signal generation}: We assume the signal of interest is being generated by multiple physical entities (e.g., TV transmitters) that are each dropped uniformly on a circle of radius $r_p$ centered at the origin. There are a total of $N_p$ such signal generators that are located at points $\{(r_{p},\theta_{j})\}_{j=1}^{N_p}$ where $\theta_{j}\sim U[0,2\pi],\bsp \forall j$. The vector signal emitted at time $t$ is denoted by ${\bf s}(t) = [s_1(t)\bsp s_2(t)\ldots s_{N_p}(t)]^T$ where $s_m(t)$ corresponds to physical entity or signal generator $m$ and comes from some finite, discrete set.

\begin{figure}[h]
\centering
\includegraphics[scale=0.45]{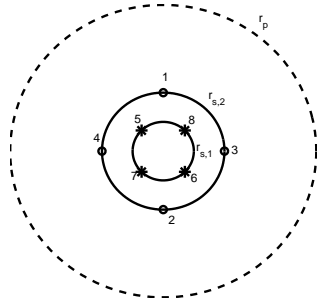}
\caption{Network with signal generators (not shown) uniformly distributed on the blue circle of radius $r_p$. There are $N_s=8$ sensing nodes in the network equally-divided across two circles ($q=2$) of radii $r_{s,1}$ and $r_{s,2}$ respectively. This gives rise to partitions ${\mathcal C}_1=\{1,2,3,4\}$ and ${\mathcal C}_2=\{5,6,7,8\}$. The sensing nodes are equally-spaced on each circle as shown.}
\label{sysmodel}
\end{figure}

{\bf Spatial distribution of sensing nodes}: There are $N_s,\bsp N_s\geq N_p$ sensing nodes that are placed on a collection of $n_c$ circles of radii $\{r_{s,1},r_{s,2},\ldots, r_{s,n_c}\}$ where $N_s$ is such that $\frac{N_s}{n_c}$ is \textit{even}. Circle $c$ contains $\frac{N_s}{n_c}$ interfering receivers located at fixed points $\{(r_{s,c},\theta_i)\}_{i=1}^{N_s}$ that are equally-spaced $\left(\theta_i=\frac{2\pi n_c}{N_s},\bsp i=0,1,\ldots,\frac{N_s}{n_c}-1\right)$ as shown in Fig.~\ref{sysmodel}. We note that this would roughly be the case when $N_s$ becomes large and the users are uniformly distributed on this collection of circles. The fusion center is located at some arbitrary point on the $xy$-plane.

The above spatial distribution model affords us analytical tractability. In the simulations section, we show that the proposed algorithms work even under a more general spatial model where both the sensing and signal nodes are scattered uniformly on a regular square grid. For the sake of the analysis, we will also partition the sensing nodes according to the circle they belong to thus creating $n_c$ partitions $\{{\mathcal C}_1,{\mathcal C}_2,\ldots,{\mathcal C}_{n_c}\}$ such that $\bigcup_{i=1}^{n_c}{\mathcal C}_i=\{1,2,\ldots,N_s\}$ and ${\mathcal C}_i\cap {\mathcal C}_j=\emptyset$ for $i\neq j$. Within each circle, the nodes are numbered or ordered in diametrically opposite pairs as shown in Fig.\ref{sysmodel}, a labelling rule that is feasible since $\frac{N_s}{n_c}$ is even. In other words, all pairs $(j,j+1)\in{\mathcal C}_i,\bsp j\bsp odd$, will correspond to a pair of diametrically opposite nodes on circle ${\mathcal C}_i$.

{\bf Channel model:} The distance between the $n$-th signal generator at location $(r_p,\theta_n)$ and $m$-th sensing node at $(r_m,\theta_m)$ is given by
$$
d_{mn}=\sqrt{r_m^2+r_n^2-2 r_m r_n\mbox{cos}(\theta_m-\theta_n)},
$$
which induces the following path-loss gain
\be
\lambda_{mn}=\frac{1}{K+d_{mn}^2},\bsp K>0.\label{PLdefinition1}
\ee
between the same. This model is an approximation of the free space path-loss model (with path-loss coefficient two) \cite{rappaport}, a choice that affords us analytical tractability while compromising very little on modeling accuracy. The composite channel gain between $n$-th signal generator at location $(r_n,\theta_n)$ and $m$-th sensing node at $(r_p,\theta_m)$ is given by
\be
h_{mn}(t)=g_{mn}(t)\sqrt{\lambda_{mn}},\label{comp_channel_definition}
\ee
where $g_{mn}(t)\sim {\mathcal CN}(0,1)$ models small-scale Rayleigh fading. Distances do not change as a function of time since sensor nodes are deterministically placed and signal nodes are assumed fixed, once drawn from a uniform distribution.

{\bf Sensor measurement model:} We adopt the sensing model in \cite{peh,sahai}, where each sensor collects a set of samples over a window of size $W$ time slots. In particular, at time $t$ sensor node $m$ receives the signal
$$
r_m(t) = \sum_n h_{nm}(t)s_n(t) + v_m(t),
$$
where $v_m(t)$ is additive noise. Each sensor then forms the following \emph{measurement} over a $W$-window of samples,
\be
z_n(k) = \frac{1}{W}\sum_{t = kW+1}^{(k+1)W} |r_n(t)|^2 - {\mathbb E}[|v_n(t)|^2].\label{measurement_raw}
\ee
The sensors then send their measurements to the fusion center via orthogonal error-free control channels. The fusion center is charged with the task of recovering the observed signal $\{s_n(t)\}$. In this work, we do not explicitly account for the time incurred in transmitting measurements to the fusion center and for subsequent actions such as data transmissions by the secondary network in the case of cognitive radio, as we are primarily interested in reducing the control bandwidth between the sensors and the fusion center. Thus, the time index $t$ (and hence $k$) iterates only across sensing windows.

{\bf Sparse dynamics:} Due to mismatches between the time-scale of the signal generators and the sensor sample collection period, we assume that the vector ${\bf s}(t)$ exhibits the following behavior. The signal $s_m(t)$ remains constant through the $k$-th collection window $kW+1\leq t < (k+1)W$ and is denoted by $s_m(k)$. Furthermore, only a subset of signals change between the $k$-th and the $(k+1)$-th collection window, i.e., $|{\bf s}(k+1)-{\bf s}(k)|$ is a sparse vector. This model is effective when the signal values change on a slower time-scale in relation to the sensing timeline. This is indeed what one typically expects for systems that exhibit a time-scale separation between the sensing network and the signal emitting process. Under the sparse dynamics assumption and for $W$ sufficiently large, we invoke the \emph{Law of Large Numbers (LLN)} and write the measurements in (\ref{measurement_raw}) as
\be
{\bf z}(k) = \boldsymbol{\Lambda}{\bf s}(k) + \boldsymbol{\beta}(k),\label{measurement}
\ee
where $\lambda_{mn} = {\mathbb E}[|h_{mn}(t)|^2]$ from (\ref{comp_channel_definition}); $\beta_n(k)$ is a noise term that models the inaccuracies of the LLN over finite averaging windows. It is well-known from the \emph{Central Limit Theorem} that ${\mathbb E}[|\beta_n(k)|^2]=O\left(\frac{1}{N}\right),\bsp \forall n$. The term $\boldsymbol{\beta}(k)$ is referred to as LLN-noise through the remainder of this paper. Next, in the context of the measurement model in (\ref{measurement}), we introduce the recovery algorithm that will be developed further through the course of this paper to exploit the sparse structure of the observed dynamics.

{\bf Recovery algorithm:} We assume that a subset $Q\subseteq\{1,2,\ldots,N_s\}$ of the measurements ${\bf z}_{Q}(k)$ are transmitted through $|Q|$ orthogonal control channels to the fusion center at the end of sensing window $k$. The fusion center must recover the signal ${\bf s}(k)$. The na\"{i}ve approach to recovering ${\bf s}(k)$ would be to transmit \emph{all} the measured values, i.e., set $Q = \{1,2,\ldots,N_s\}$ to the fusion center\footnote{The fusion center may recover the signal using standard tools such as least-squares.}. The above na\"{i}ve solution would consume a control bandwidth of ${\mathcal O}(N_s)$. If the process ${\bf s}(k)$ is completely general, there is little that can be done to remedy this problem, and partial feedback (of only a subset of ${\bf z}$) will necessarily result in degraded performance. However, as discussed in the introduction, for networks where only a subset of signal generators (e.g., TV transmitters) change their state between two adjacent sensing windows, we show that it is possible to reduce the control bandwidth. Using ideas from subset selection and compressive sampling, the next section considers how sparse dynamics can be exploited in order to achieve near-optimal performance while reducing the bandwidth consumption.

\section{Recovering dynamics through compressive sampling}\label{sec: CS}
In this section, we propose a compressive sampling approach to efficiently recover the signal ${\bf s}(k)$. Before we discuss the recovery algorithm, we take a short diversion into the topic of compressive sampling.

\subsection{Compressive sampling}
The topic of compressive sampling has received tremendous interest in the recent years \cite{candesSigProcMag,baraniuk,donoho1}. The theory essentially states that one can recover sparse data exactly, given an under-determined system of equations. Specifically, the generic problem is the following: Given a signal ${\bf x} \in \mathbb{R}^N$, one receives $q<<N$ linear, potentially noisy measurements: $\mathbf{y} = {\bf M} {\bf x} + \mathbf{w}$. Here, ${\bf M}\in{\mathbb R}^{q\times N}$ encodes the measurement matrix and ${\bf w}\in{\mathbb R}^{q}$ denotes additive noise, usually of bounded norm.

For a general vector ${\bf x} \in \mathbb{R}^N$, $N$ independent measurements are required to hope to reconstruct ${\bf x}$. When $q < N$, the problem therefore is under-determined. If ${\bf x}$ is sparse, however, in some settings the problem is no longer under-determined, and can be solved exactly by considering standard convex programs such as
$$
\arg \min_{{\bf x}\in{\mathbb R}^p}:||{\bf M} {\bf x} - \mathbf{y}||_2^2 - \xi ||{\bf x}||_1\mbox{ for some chosen }\xi > 0,
$$
in the noisy case and
\be
\begin{array}{rcl}
\arg \min: && ||\bf x||_1 \\
&& \bf M \bf x = \mathbf{y}.\label{Ourl1min}
\end{array}
\ee
in the noiseless case. The former is the so-called \emph{Lasso} \cite{lasso} formulation for model selection (subset selection) while the latter $\ell_1$-norm minimization problem is called \emph{Basis Pursuit}. Many such results have appeared in the literature, e.g., \cite{candesRIP,candesSigProcMag,baraniuk,donoho1} under the umbrella of compressive sampling\footnote{We do not use the terminology ``compressed sensing'' to avoid confusion with ``cooperative sensing''.}. Indeed, the results are attractive from an algorithmic perspective since the convex relaxation is easily solvable, with computation time that scales gracefully as the size of the problem increases, allowing the efficient solution of very large problems. The above convex programs succeed as long as the linear equations, or measurements, satisfy a property called \textit{Null Space Property} (NSP), which essentially amount to the statement that there are no very sparse vectors in the null-space of the measurement matrix $\bf M$. The theoretical connections between Lasso and Basis Pursuit have been well-analyzed by authors such as Tropp \cite{troppLASSO}.

\subsection{The recovery algorithm}

Returning to our problem, we define $\Delta {\bf s}(k)={\bf s}(k) - {\bf s}(k-1)$ and apply the model selection paradigm outlined above, to {\em the dynamics vector} $\Delta {\bf s}(k)$ rather than the signal vector itself. We can assume that at some initial time $k_0$, ${\bf s}(k_0)$ is known. At time $k$, we ``query'' a subset of sensors ${\mathcal Q}$ and receive measurements $\mathbf{z}_{{\mathcal Q}}(k) = \boldsymbol{\Lambda}_{\mathcal Q}{\bf s}(k) + \boldsymbol{\beta}(k)$. We can then construct the difference in measurements
\begin{eqnarray*}
\Delta\mathbf{z}_{{\mathcal Q}}(k) &=& \mathbf{z}_{{\mathcal Q}}(k) - \mathbf{z}_{{\mathcal Q}}(k-1)\\
&=& \boldsymbol{\Lambda}_{\mathcal Q} [\mathbf{s}(k) - \mathbf{s}(k-1)] + \boldsymbol{\beta}(k) - \boldsymbol{\beta}(k-1)\\
&=&  \boldsymbol{\Lambda}_{{\mathcal Q}} \Delta \mathbf{s}(k) + \boldsymbol{\beta}(k) - \boldsymbol{\beta}(k-1).
\end{eqnarray*}
Since the left hand side, $\Delta\mathbf{z}_{{\mathcal Q}}(k)$, is known, this falls precisely into the sparse recovery paradigm developed above. More concretely, let $Q$ be the subset of queried users with size $|Q| = q,\bsp q\bsp even$. Then, $Q$ is chosen according to the following algorithm:\\
\begin{algorithm}
\caption{Protocol to choose query set $Q$ of size $|Q| = q$} \label{alg: feedback}
\begin{algorithmic}[1]
\STATE Set $Q=\emptyset$ and $i = 0$.

\WHILE{$i+1\leq \frac{q}{2}$}

\STATE Choose any pair of diametrically opposite receivers $(j,j+1)$ from circle $i+1$, i.e., $j\in{\mathcal C}_{i+1},\bsp j\bsp odd$.

\item $Q = Q\cup \{j,j+1\}$.

\item Set ${\mathcal C}_{i+1}={\mathcal C}_{i+1}\setminus \{j,j+1\}$

\STATE Increment $i=(i+1) (\mbox{mod }n_c)$.

\ENDWHILE
\end{algorithmic}
\end{algorithm}
Note that the output query set is not unique since the choice of node-pairs is left open. Let ${\mathcal Q}$ be the set of all possible output query sets from Algorithm~\ref{alg: feedback} above. For the example in Fig.~\ref{sysmodel}, one possible output query set for $q = 4$ is $Q=\{1,2,7,8\}$. It is necessary for a query set $Q$ to be selected in this way for the sake of analytical tractability. As with the spatial distribution model, we adopt a more general querying model in our simulations section. Under such a querying model, it is of immediate interest to determine the smallest query size $q$ (or control bandwidth) that the system requires in order to recover ${\bf s}(k)$ reliably using
\begin{eqnarray}
&\arg\min_{{\bf x}\in{\mathbb R}^{N_p}}: || \mathbf{\Lambda}_{{\mathcal Q}}{\bf x} - \Delta\mathbf{z}_{{\mathcal Q}}(k)||_2^2 - \xi ||{\bf x}||_1.\label{lasso}
\end{eqnarray}
As mentioned earlier, in this work, we do not consider the number of bits required to communicate ${\bf z}_{Q}(k)$ reliably to the fusion center as we are interested primarily in the scaling behavior of the control bandwidth.

Compressive sampling theory states that recovery of any $S$-sparse vector through Lasso or Basis Pursuit is possible in a noiseless setting if and only if the \textit{sensing matrix} ${\boldsymbol{\Lambda}}_{Q},\bsp Q\in{\mathcal Q}$ satisfies the NSP \cite{cohen} of order $S$. This property is defined in the next section. We note that in our application, the sensing matrix is provided by the channel as opposed to traditional compressive sampling where the designer is allowed to choose a convenient sensing mechanism. In the next section, we present the main result of this paper, which states path-loss matrices ${\boldsymbol{\Lambda}}_{Q},\bsp Q\in{\mathcal Q}$ make for good sensing matrices in the noiseless case, i.e., when $W\rightarrow \infty$. Recovery results that are proved for Basis Pursuit, which, in the absence of noise, carry over to Lasso \cite{troppLASSO}. While the theory is developed for the noiseless case with $W\rightarrow \infty$, our simulations consider finite averaging windows and demonstrate that sparse recovery still remains effective in this setting.

\section{NSP of path-loss matrices}\label{sec: NSPPL}

In this section, we establish that path-loss matrices ${\boldsymbol{\Lambda}}_{Q},\bsp Q\in{\mathcal Q}$ satisfy the \emph{Null Space Property} (which will be defined shortly) when the control bandwidth obeys $q={\mathcal O}(S\mbox{log }N_p)$. Lemma~\ref{mainlemma1}, Lemma~\ref{mainlemma2} along with Theorem~\ref{mainresult} constitute the main results in this section.

\subsection{Preliminaries}
Let the support set of ${\bf x}$ be denoted by ${\mathcal S}$. A vector ${\bf x}$ is $S$-sparse if $|{\mathcal S}|\leq S$. We define the null space property from Gribonval et al. \cite{gribonval}. Given a matrix ${\bf M}$, let ${\mathcal N}({\bf M})$ denote its null space.

\noi \textit{Definition (Null space Property)}: A matrix ${\bf M}$ satisfies the null space property of order $S$ if for all subsets ${\mathcal S}\subseteq \{1,2,\ldots,N\}$ with $|{\mathcal S}|\leq S$, the following holds
\ben
||{\bf v}_{{\mathcal S}}||_1\leq ||{\bf v}_{{\mathcal S}^c}||_1,\bsp \forall {\bf v}\in{\mathcal N}({\bf M})\setminus {\bf 0}.
\een
where ${\mathcal S}^c= \{1,2,\ldots,N\}\setminus {\mathcal S}$. Based on this property, the following recovery result \cite{gribonval} has appeared both implicitly and explicitly in works such as \cite{donoho3,cohen}.
\begin{theorem}\cite{gribonval}\label{theoremNSP}
Let ${\bf M}\in{\mathbb R}^{q\times N}$. Then, any $S$-sparse vector may be recovered by solving (\ref{Ourl1min}) iff ${\bf M}$ satisfies the NSP of order $S$.
\end{theorem}$\hfill\Box$\\
The NSP is typically quite difficult to prove directly leading to the development of sufficient conditions that are easier to establish. One such sufficient condition is the \textit{restricted isometry property} \cite{candesRIP} that has become quite popular in recent years and is defined below.

\noi \textit{Definition (Restricted Isometry Property)}: A $q\times N$ matrix ${\bf M}$ satisfies the Restricted Isometry Property (RIP) of order $p$ if there exists $\epsilon_p({\bf M})\in(0,1)$ such that for all ${\bf x}\in {\mathbb R}^N$,
\be
\bs
(1-\epsilon_p({\bf M}))||{\bf x}_{{\mathcal T}}||^2_2\leq ||{\bf x}_{{\mathcal T}}^T({\bf M}^T)_{{\mathcal T}}||_2^2\leq (1+\epsilon_p({\bf M}))||{\bf x}_{\mathcal T}||^2_2
\es
\ee
holds for all sets ${\mathcal T}$ with $|{\mathcal T}|\leq p$. Here, $({\bf M}^T)_{{\mathcal T}}$ is the sub-matrix of ${\bf M}^T$ formed by rows in ${\mathcal T}$.

Here, $\epsilon_{p}({\bf M})$ is called the \textit{restricted isometric constant} of ${\bf M}$. The RIP essentially requires that all $q\times |{\mathcal T}|$ sub-matrices of ${\bf M}$ be well-conditioned. Under such a conditioning, perfect recovery of ${\bf x}$ is possible as stated in the following theorem.
\begin{theorem}\cite{GribonvalDavies,lai}\label{theoremRIP}
Let ${\bf M}\in{\mathbb R}^{q\times N}$. If  ${\bf M}$ satisfies the RIP with $\epsilon_{2S}({\bf M})\leq 2\frac{(3 -\sqrt{2})}{7}\approx 0.4531$, then every $S$-sparse vector ${\bf x}\in{\mathbb R}^N$ is the solution to the $\ell_1$-norm minimization problem in (\ref{Ourl1min}).
\end{theorem}
$\hfill\Box$\\

Thus, the RIP with a sufficiently small constant immediately implies the NSP in the context of $\ell_1$-recovery. The approach we use to prove ``goodness'' of path-loss matrices ${\boldsymbol{\Lambda}}_{Q}$ is motivated by the following observation. In general, the null space of a product of two matrices ${\bf N}{\bf M}$ contains the null space of ${\bf M}$ and therefore if ${\bf N}{\bf M}$ satisfies the NSP, so does ${\bf M}$. This allows us to study the class of \textit{linearly-processed} path-loss matrices ${\bf A}_{Q}={\bf B}\underbrace{{\bf W}{\boldsymbol{\Lambda}}_{Q}}_{{\bf G}}$ where
\be
{\bf W}=\mbox{diag}\{\underbrace{{\bf J}\bsp {\bf J}\ldots {\bf J}}_{\frac{q}{2}\mbox{ times}}\},\bsp {\bf J}=\left[\begin{array}{rr}
1&-1\\
-1&1\ea\right],\label{Gdef}
\ee
and $\mbox{diag}\{\cdot\}$ is the standard block-diagonal operator;
\be
{\bf B}=\mbox{diag}\left\{\frac{\beta_{1}}{\sqrt{\mbox{Var}\{g_{11}\}}}\ldots \frac{\beta_{q}}{\sqrt{\mbox{Var}\{g_{q1}\}}}\right\}
\ee

with $\beta_{i}\sim \mbox{Bernoulli}\left(\frac{1}{2}\right),\bsp \forall i$ and independent across $i$. The Bernoulli random variables have support $\{\pm 1\}$. We focus our attention on establishing the recovery properties of ${\bf A}_{Q}$ rather than ${\boldsymbol{\Lambda}}_{Q}$. We show that ${\bf A}_{Q}$ satisfies the RIP with $q={\mathcal O}(S\mbox{log }N_p)$ measurements and hence the NSP. The transformation ${\bf W}$ essentially subtracts rows of ${\boldsymbol{\Lambda}}_{Q}$ corresponding to \textit{diametrically opposite} pairs of sensors in the \textit{same} partition. Thus, the dimension of ${\bf G}$ is still $q\times q$. The transformation ${\bf B}$ weights and adds adjacent rows of ${\bf G}$.

According to our spatial distribution model, since the positions of the sensing nodes are fixed, the columns of ${\boldsymbol{\Lambda}}_{Q}$ become stochastically independent since each interfering transmitter is independently thrown. At this point, we will rely heavily on recent results from Vershyin \cite{vershynin} and Adamcyzk et al. \cite{adamcyzk} that deal with sensing matrices containing independent columns. Before we reproduce the RIP result \cite{vershynin,adamcyzk} for matrices with independent columns, we present a primer on sub-gaussian and sub-exponential random variables along with some useful results from non-asymptotic matrix theory.

\subsection{Useful concentration inequalities}

We refer the reader to the tutorial paper by Vershynin \cite{vershynin} for a great introduction to non-asymptotic matrix theory. Lemmas~\ref{lemmasubgaussian}-\ref{lemmasubgaussianvectors} below are well-known past results that are summarized in this paper \cite{vershynin}. The proofs are not reproduced due to lack of space.
\begin{lemma}\label{lemmasubgaussian}
Let $u$ be random variable. The following properties are equivalent with parameters $K_i > 0,\bsp i  = 1,2,3,4$, differing from each other by at most an absolute constant factor.\\
\noi {\em (i)} Tails: $\mbox{Pr}(|u|>\gamma)\leq \exp(1 -\frac{\gamma^2}{K_2})$ for all $\gamma>0$, \\
\noi {\em (ii)} Moments: $({\mathbb E}\left[|u|^p\right])^{\frac{1}{p}} \leq K_2\sqrt{p}$ for all $p\geq 1$,\\
\noi {\em (iii)} Super-exponential moment: ${\mathbb E}\left[\mbox{exp}\left(\frac{u^2}{K_3}\right)\right]\leq \exp(1)$.\\
Moreover, if ${\mathbb E}[u]= 0$ then properties \textit{(i)}-\textit{(iii)} are also equivalent to the following property:\\
\noi \textit{(iv)} Moment generating function: ${\mathbb E}\left[\mbox{exp}\left(u\gamma\right)\right]\leq \mbox{exp}(\gamma^2K_4)$ for all $\gamma\in{\mathbb R}$.
\end{lemma}$\hfill\Box$\\
A random variable that satisfies the above property is called a \textit{sub-gaussian} random variable. Such random variables are often characterized by the $\psi_2$-norm\footnote{Alternate definitions of this norm have been adopted (such as in \cite{adamcyzk}) that are all equivalent to within a constant factor.}, which is defined as
\ben
||u||_{\psi_2}= \mbox{sup}_{p\geq 1} \frac{({\mathbb E}\left[|u|^p\right])^{\frac{1}{p}}}{\sqrt{p}}.
\een
It follows that if the $\psi_2$-norm of $u$ is finite, then $u$ is a sub-gaussian random variable with $||u||_{\psi_2}=K_2$. This is in fact the case for bounded random variables with symmetric distributions.
\begin{lemma}\label{lemmasymrvs}
Let $u$ be a symmetrically distributed, bounded random variable with $|u|\leq M,\bsp M>0$. Then, $u$ is a sub-gaussian random variable with $||u||_{\psi_2}\leq c M^2,\bsp c > 0$.
\end{lemma}\hfill$\Box$

In higher dimensions, a random vector ${\bf u}$ of dimension $N$ is called sub-gaussian if ${\bf u}^T{\bf x}$ is sub-gaussian for every ${\bf x}\in{\mathbb R}^N$.

\begin{lemma}\label{lemmasubgaussianvectors}
Let $\{u_i\}_{i=1}^p$ be a collection of independent, zero-mean, sub-gaussian random variables. Then, ${\bf u}$ is a sub-gaussian random vector with $||{\bf u}||_{\psi_2}=C\max_{i}\bsp ||u_i||_{\psi_2}$ for some $C>0$.
\end{lemma}\hfill$\Box$\\
We are now ready to prove the RIP (hence NSP) for matrix ${\bf A}_{Q}$. Before we move on to this task, we require one more definition. A random vector ${\bf m}$ of dimension $N$ is called \textit{isotropic} if ${\mathbb E}[|{\bf m}^T{\bf x}|^2]=||{\bf x}||^2$ for all ${\bf x}\in {\mathbb R}^N$.

\subsection{NSP of linearly-processed path-loss matrices ${\bf A}_{Q}$}

We reproduce the recent RIP result \cite{vershynin,adamcyzk} concerning matrices with independent columns. We refer the reader to \cite{vershynin,adamcyzk} for the proof.
\begin{theorem}\label{RIPindpcolumns}
Let ${\bf M}=\left[{\bf m}_1\bsp {\bf m}_2\ldots{\bf m}_N\right]$ be a $q\times N$ random matrix whose columns are independent, isotropic and sub­-gaussian with $\psi_{max,m}=\max_{i\in\{1,\ldots,N\}}\bsp  ||{\bf m}_i||_{\psi_2}$. Furthermore, let the columns satisfy $||{\bf m}_i||^2=q$ almost surely for all $i$. Then, the normalized matrix $\frac{1}{\sqrt{q}}{\bf M}$ is such that if $q\geq C_{\psi_{max,m}}\varepsilon^{-2}S\mbox{log}\left(\frac{\exp(1)N}{S}\right)$ for some $\varepsilon>0$, then $\varepsilon_p\left(\frac{1}{\sqrt{q}}{\bf M}\right)\leq \varepsilon$ with probability at least $1-2\exp(-c_{\psi_{max,m}}\varepsilon^2 q)$. Here, $c_{\psi_{max,m}}$ and $C_{\psi_{max,m}}$ are positive scalars that depend only the worst-case sub-gaussian norm $\psi_{max,m}$.
\end{theorem}
\hfill$\Box$\\
As mentioned earlier, the channel matrix ${\boldsymbol{\Lambda}} = \left[\boldsymbol{\lambda}_1\bsp \boldsymbol{\lambda}_2\ldots \boldsymbol{\lambda}_{N_p}\right]$ contains independent columns since the positions of the sensing nodes are fixed. However, each column contains entries that are not centered, not isotropic and that are highly coupled. This is because all entries in ${\boldsymbol{\lambda}}_i$ are now completely determined by the position of signal generator $i$. For this reason, it is not immediately clear whether the columns are sub-gaussian.

To finally prove the NSP of ${\boldsymbol{\Lambda}}_{Q},\bsp Q\in{\mathcal Q}$, we show that after
left-multiplication by matrices $\bf B$ and $\bf G$, the matrix ${\bf A}_{Q}$ satisfies the sufficient conditions in Theorem~\ref{RIPindpcolumns}. The following lemmas and theorem constitute the main theoretical results of this paper.
\begin{lemma}\label{mainlemma1}
The matrix ${\bf A}_{Q}={\bf B}{\bf W}{\boldsymbol{\Lambda}}_{Q},\bsp Q\in{\mathcal Q}$ of size $q\times N$ contains independent, isotropic, centered, sub-gaussian columns.
\end{lemma}
\noi {\em Proof}: See Appendix~\ref{mainproof1}. $\hfill\Box$\\
\begin{lemma}\label{mainlemma2}
For matrix ${\bf A}_{Q}={\bf B}{\bf W}{\boldsymbol{\Lambda}}_{Q},\bsp Q\in{\mathcal Q}$ of size $q\times N$, we have that $||{\bf a}_i||^2=q$ almost surely.
\end{lemma}
\noi {\em Proof}: See Appendix~\ref{mainproof2}. $\hfill\Box$\\

\begin{theorem}\label{mainresult}
${\boldsymbol{\Lambda}}_{Q},\bsp Q\in{\mathcal Q}$ satisfies NSP of order $S$ almost surely when $q={\mathcal O}(S\mbox{log}N_p)$.
\end{theorem}
\noi {\em Proof}: The result follows from Lemma~\ref{mainlemma1}, Lemma~\ref{mainlemma2} and Theorem~\ref{RIPindpcolumns}.$\hfill\Box$\\

We discuss in the next section, how this sparse recovery algorithm is incorporated into the development of a complete cooperative sensing algorithm. In particular, since we are estimating dynamics rather than the signal itself and due to the presence of LLN-noise, the real possibility of {\it error propagation} arises. This has not been heretofore addressed in the literature, to the best of our knowledge. In the next section, the algorithm introduces explicit steps to control this. The simulations in Section \ref{sec: simu} demonstrate the effectiveness of our approach.

\section{Complete cooperative sensing algorithm}\label{sec: coopsensing}

The essential conclusion of the previous section is that the signal vector ${\bf s}(k)$ can be recovered by acquiring ${\mathcal O}(S\mbox{log}N_p)$ noise-free measurements if the dynamics of ${\bf s}(k)$ are sparse.

The complete sensing algorithm is presented in Algorithm \ref{alg: 3}. Recall that we do not consider the actions taken post-sensing and the time taken (equiv. bandwidth \emph{per} control channel) to transmit the measurements to the fusion center as we are only interested in scaling properties in this work. The algorithm operates in three modes; in \emph{dynamics} mode, the algorithm retrieves signals dynamics incurring logarithmic control bandwidth and then estimates the recovery error. If the estimated recovery error is too high, the algorithm switches to \emph{partial reset} mode in the next sensing window or slot where the fusion center retrieves the signal directly while incurring linear bandwidth. Following partial reset, the system returns to dynamics mode. This operation is overlaid by a periodic (with a pre-set period) \emph{reset} mode where again the fusion center again retrieves the signal directly incurring linear bandwidth.
\begin{algorithm}[h!]
\renewcommand{\baselinestretch}{0.9}
\begin{small}
\caption{Sensing algorithm}\label{alg: 3}
\begin{algorithmic}[1]
\STATE Primary inputs: sensing window sizes $W$, number of sensors $N_s$, number of signal emitters $N_p$, sparsity factor $S$, error reset period $T_{err} > 0$
\STATE Initialize mode[0] = ``dynamics''

\FOR{each $k = 1,2,\ldots,$}

\IF {$k (\mbox{mod }T_{err}) \not\equiv 0$}

\STATE During time slots $kW\leq t + 1< (k+1)W$, each sensor samples the signal of interest $W$ times and averages the same to obtain $z_n(k)$ as per (\ref{measurement}).

\IF {mode[k-1] = ``dynamics''}

\STATE The fusion center queries $q = \Theta(S\log N_p)$ sensors indexed by set ${\mathcal Q}$ according to Algorithm~\ref{alg: feedback} and computes the change in received signal levels $\{\Delta\mathbf{z}_{Q}(k)\}.$

\STATE \emph{Dynamics mode}: Once acquired, the fusion center estimates the change in measurements $\Delta{\bf s}(k) = {\bf s}(k) - {\bf s}(k-1)$ by solving the following optimization problem:
\be
{\bf u}(k) = \mbox{\hspace{-0.1cm}}\arg\min_{{\bf x}\in{\mathbb R}^{N_p}}: || \mathbf{\Lambda}_{Q}{\bf x} - \Delta\mathbf{z}_{{\mathcal Q}}(k)||_2^2 - \xi ||{\bf x}||_1,\label{lasso_diff}
\ee
and setting $$\Delta\tilde{\mathbf{s}}(k) = \left[\phi^{\nu}_1({\bf u}(k))\bsp \phi^{\nu}_2({\bf u}(k))\bsp \phi^{\nu}_{N_p}({\bf u}(k))\right]^T,$$ where $\phi^{\nu}_i:{\mathbb R}^{N_p}\rightarrow {\mathbb R}$ is a suitably-chosen thresholding function with $\nu$ that is used to recover discrete sparse signals in noisy environments.\label{step_difference}

\STATE The fusion center sets $\tilde{\mathbf{s}}(k)=\tilde{\mathbf{s}}(k-1)+\Delta\tilde{\mathbf{s}}(k).$\label{recover_diff}

\STATE The fusion center then queries and acquires extra measurements from ${q}^{\prime}$ additional sensors $Q^{\prime}\subseteq \{1,2,\ldots,N_s\}\setminus {\mathcal Q},\bsp |{\mathcal Q}^{\prime}| = q^{\prime}$. Then, the fusion center computes
\be
e=\frac{1}{q^{\prime}}\|\mathbf{\Lambda}_{Q^{\prime}} \tilde{\mathbf{s}}(k)-\mathbf{z}_{Q^{\prime}}(k)\|_1,\label{error_measured}
\ee
which essentially represents the average recovery error on the extra measurements.\label{error_est}

\IF {$e\leq \epsilon$, i.e., the estimate is accurate,}

\STATE Set $\tilde{\mathbf{s}}(k)$ to be the final estimate of the signal.

\STATE Set mode[k] = ``dynamics''

\ELSE

\STATE Set mode[k] = ``partial reset''
\ENDIF

\ELSIF {mode[k-1] = ``partial reset''}

\STATE \emph{Partial Reset Mode}: The fusion center queries \emph{all} $N_s$ sensors and uses the measurements to directly recover the final estimate $\tilde{\mathbf{s}}(k)$ by solving
\be
{\bf u}(k) = \mbox{\hspace{-0.1cm}}\arg\min_{{\bf x}\in{\mathbb R}^{N_p}}: || \mathbf{\Lambda}_{Q}{\bf x} - \mathbf{z}_{{\mathcal Q}}(k)||_2^2 - \xi ||{\bf x}||_1\label{lasso_abs}
\ee
and again, setting
$$
\tilde{\mathbf{s}}(k) = \left[\phi^{\nu}_1({\bf u}(k))\bsp \phi^{\nu}_2({\bf u}(k))\bsp \phi^{\nu}_{N_p}({\bf u}(k))\right]^T.
$$\label{step_partial_reset}
\STATE Set mode[k] = ``dynamics''

\ENDIF

\ELSIF {$k (\mbox{mod }T_{err}) \equiv 0$}

\STATE During time slots $kW\leq t < (k+1)W$, each sensor forms $z_n(k)$ as per (\ref{measurement}).

\STATE \emph{Full Reset Mode}: The fusion center then queries \emph{all} sensors and uses the $N_s$ measurements to directly recover the final estimate $\tilde{\mathbf{s}}(k)$ by solving (\ref{lasso_abs}). \label{step_reset}

\STATE Set mode[k] = ``dynamics''.

\ENDIF

\ENDFOR
\end{algorithmic}
\end{small}
\end{algorithm}

In a typical sensing slot, the Lasso procedure in Step \ref{step_difference} recovers $\Delta\tilde{{\bf s}}(k)$. The recovery is almost surely exact when $\Delta{\bf s}(k)$ is sparse and when $W$ is sufficiently large. However, as this might not always be the case in reality, error propagation is a critical aspect that needs to be managed when operating on the dynamics of the signal. The following remarks describe the steps taken by the algorithm to address the same:
\begin{itemize}

\item The system selects a parameter $T_{err}$ that represents the number of sensing slots after which the fusion center must query \emph{all} sensors and recover the absolute signal as opposed to the dynamics.

\item During typical operation, i.e., when $k(\mbox{mod }T_{err})\not\equiv 0$, we estimate the accuracy of $\tilde{\mathbf{s}}(k)$ in Step~\ref{error_est} immediately following compressed sampling in Step~\ref{recover_diff}. The technique is commonly referred to as \emph{cross-validation} \cite{ward}. If the estimated error is larger than the threshold $\epsilon$, we query \emph{all} remaining $N_s-(q+q^{\prime})$ sensors and recover the absolute signal in the next time slot.

\item Once more, due to the sources of noise described above, the vector $\mathbf{u}(k)$ retrieved by Lasso in (\ref{lasso_diff}) and (\ref{lasso_abs}) will almost never be truly sparse. Thus, we introduce a thresholding function to recover a discrete signal.

\end{itemize}

\section{Numerical results}\label{sec: simu}
In this section, we evaluate the performance of the proposed sensing algorithm using numerical experiments.

\subsection{Simulation setup}
The simulation setting is described in the following.\\

\textbf{Network geography} - We consider a square region with side length $1$ kilometer. We partition this square area into $N = 49$ micro-cells, each micro-cell is  $143m\times143m$ square area, which is sufficiently small to ensure at most one signal generator or sensor per micro-cell. We index the micro-cells by $i$ ($1\leq i\leq N$), counted column-wise as shown in Figure \ref{fig:sim_grid}. Each micro-cell contains one {\em potential} signal generator that is located at the center of the square micro-cell. The sensors are positioned one per micro-cell, i.e, $N_s = N = 49$.
\begin{figure}[h]
\centering
{\includegraphics[scale=0.35]{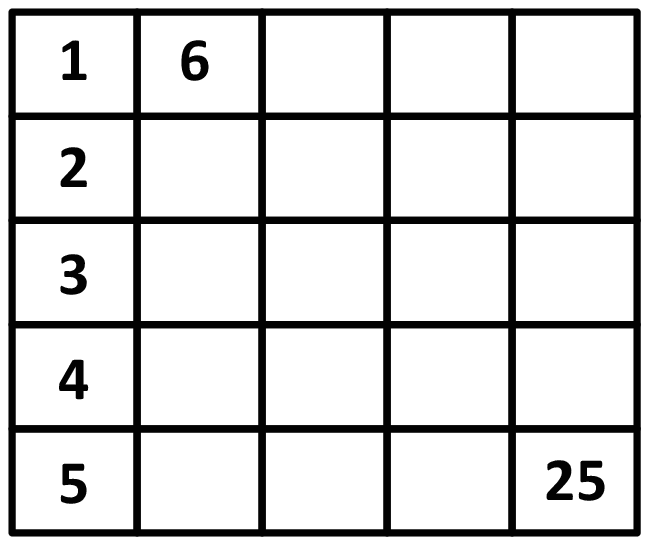}}
\caption{Illustrative example of grid model used in simulations with $N = 25$ and $N_s=4$. }
\label{fig:sim_grid}
\end{figure}

\textbf{Signal dimension and channel state knowledge} - By virtue of the above geographical model where we have ``discretized'' the locations of the sensors and signal generators, the fusion center may now pre-compute $\boldsymbol \Lambda$ based on the grid. This removes the need for real-time knowledge about the positions of the signal generators. However, this comes at a price, namely, an increase in the dimension of the signal from $N_p$ to $N$. Thus, the value $s_m(k),\bsp m =1,2,\ldots,N$, is non-zero if a signal generator is \emph{present and active} at micro-cell $m$ on the grid.

\textbf{Signal activity} - In our simulations, we assume that the signals are binary-valued, i.e., $s_m(k)\in\{0,P\},\bsp P > 0,\forall m,k$. The activity of each signal generator or micro-cell $s_m(k)$ is modeled as a Markovian ON-OFF process where the transition probability from $0$ to $P$ is $\eta_0 = \eta = 0.1$ and from $P$ to $0$ is $\eta_1 = \eta = 0.1$. The probability that the signal changes state from sensing window to the next can be computed as $\frac{2\eta_0\eta_1}{(\eta_0+\eta_1)} = \frac{2\times 0.1\times 0.1}{(0.1+0.1)}=0.1$. Thus, on average $10\%$ or roughly $S_{ave} = 49\times 0.1\approx 5$ signals change their state, which reflects the sparse dynamics. As per our system model, we assume that $P$ is known and thus, we are only interested in estimating the support at time $k$. Henceforth, the true support at time $k$ be given by ${\mathcal P}(k) = \{m:m=1,2,\ldots,N,\bsp s_m(k) = P\}$, while the estimated support by Algorithm~\ref{alg: 3} is given by $\hat{{\mathcal P}}(k)= \{m:m=1,2,\ldots,N,\bsp \hat{s}_m(k) = P\}$.

\textbf{Post-recovery thresholding} - We use a simple per-element thresholding function
\be
\phi^{\nu}_i({\bf x}) = \left\{\begin{array}{lr}
\mbox{sign}(x_i)P,&|x_i| \geq \nu\\\label{rounding_func}
0,&\mbox{else}
\end{array}\right.
\ee
to recover the actual sparse entries themselves. Note that we have to account for negative values for $x_i$ since $\mathbf{u}(k)$ represents differences in (\ref{lasso_diff}).

\textbf{Performance metric} - To measure the signal recovery accuracy of our algorithm, we use a set distance. Given two sets ${\mathcal A},{\mathcal B}\subseteq\{1,2,\ldots,N\}$, the measure of distance that we use is
\be
d({\mathcal A},{\mathcal B}) = 1-\frac{1}{N}\left[|{\mathcal A}\cap {\mathcal B}|+ |{\mathcal A}^c\cap {\mathcal B}^c|\right].\label{distance_metric}
\ee
In the context of our sensing problem, i.e., setting ${\mathcal A} = {\mathcal P}(k)$ and ${\mathcal B} = \hat{{\mathcal P}}(k)$, we see that $d(\cdot)$ essentially counts the fraction of elements that are incorrectly recovered at time $k$. Now since we are dealing with a stochastic, time-varying system, we are interested in the long-term average distance that is given by
\be
\frac{1}{T}\sum_{k = 1}^T d({\mathcal P}(k),\hat{{\mathcal P}}(k)),\label{ave_distance}
\ee
where $T$ is the length of our simulation counted as the total number of sensing windows.

\textbf{Other algorithm parameters} - The number of measurements for recovery and error measurement are set to $q = S_{ave}\log N = \frac{2\eta^2}{(\eta_0+\eta_1)} \log N = \eta \log N  = 29$ and $q^{\prime} = 5$ with associated query sets $Q=\{1,2,\ldots,k\}$ and $Q^{\prime}=\{k+1,k+q^{\prime}\}$. The regularization parameter is set to $\xi = 0.01$. The remaining parameters such as $W$,$\eta$,$\varepsilon$ and $T_{err}$ will be varied in our experiments.

 \textbf{Control bandwidth} - Note that the parameters $q$, $q^{\prime}$, $\varepsilon$, $T_{err}$  and $N$ have a direct bearing on the bandwidth consumed by our algorithm. To analyze the bandwidth consumed in Algorithm~\ref{alg: 3}, we basically count the fraction of time algorithm spends in each of the modes in Steps~\ref{step_difference}, \ref{step_partial_reset} and \ref{step_reset}. For a total communication duration of $T$ sensing windows or slots, i.e., $k = 1,2,\ldots,T$, let the time spent in dynamics, semi-reset and reset mode be given by $T_{d}$, $T_{s-r}$ and $T_r$ respectively. Then, the bandwidth consumed, is given by
\be
\begin{small}
\ba
B &=& \frac{T_{d}}{T} (q + q^{\prime}) + N\left(\frac{T_{s-r}+T_{r}}{T}\right)\\
 &=& \frac{T_{d}}{T} (q + q^{\prime}) + N\left(1 - \frac{T_{d}}{T}\right)\label{bw_formula}
 \ea
\end{small}
\ee
and the savings over full (linear) control bandwidth can be expressed as $1 -\frac{B}{N}$. The control bandwidth, in addition to the distance measure in (\ref{distance_metric}), is an important evaluation metric, one that forms the core motivation of this paper.

Through the remainder of this section, we explore the many interesting trade-offs that are inherent in our proposed sensing design in the context of the above parameters.

\subsection{Simulation results}
Having described our simulation setup in detail, we now proceed to analyze the performance of our algorithm. Since the baseline na\"{i}ve algorithm in itself does not achieve perfect recovery due to LLN noise, we adopt the following methodology for comparison. We fix a target average distance and determine the minimum sensing window size $W_{min}$ needed to achieve this target when consuming full linear bandwidth, i.e., with $q = N$. Note that as the sensing window size grows, the average distance can be made arbitrarily small thereby approaching perfect recovery. We show that with careful tuning, Algorithm~\ref{alg: 3} consumes considerably lesser bandwidth in achieving the same target average distance while maintaining the same window size of $W_{min}$.

First, in order to determine $W_{min}$, we analyze the average distance of the na\"{i}ve algorithm versus $W$, i.e, set $T_{err} = 1$ in Algorithm~\ref{alg: 3}, which means that we always retrieves absolutes. We fix the target average distance to be $0.1$. In Fig.~\ref{distanceVsW}, we plot the average distance of the na\"{i}ve algorithm over the range $W\in\{50,\ldots,1000\}$. From Fig.~\ref{distanceVsW}, we see that for a target average distance of $0.1$, we need $2000$ samples to form (\ref{measurement}). We therefore set $W_{min} = 2000$ as the baseline window size for the na\"{i}ve algorithm.
\begin{figure}[hbt]
\centering
{\includegraphics[scale=0.6]{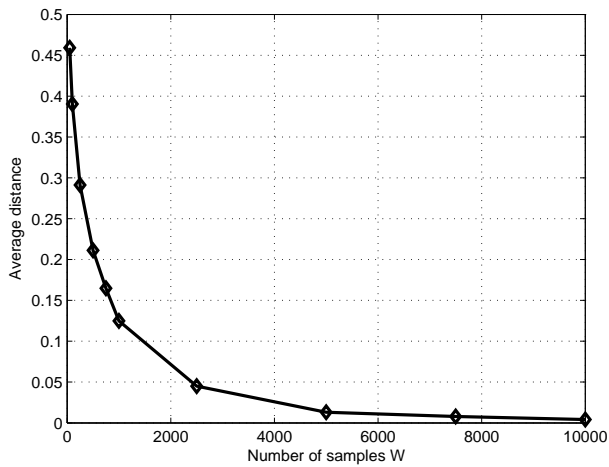}}
\caption{Average distance of the full linear bandwidth algorithm versus $W$, i.e, $T_{err} = 1.$}
\label{distanceVsW}
\end{figure}

Having identified $W_{min}$, next we proceed to tune Algorithm~\ref{alg: 3} with the goal of outperforming the na\"{i}ve algorithm. In particular, we study the accuracy-bandwidth performance of our approach while varying the error threshold $\varepsilon$, the reset period $T_{err}$ and the support threshold $\nu$ over the space $(\varepsilon,T_{err},\nu)\in\{0.1P,0.5P,P,2P,3P\}\times \{P,2P,3P,4P,5P,8P\} \times \{0.25P,0.5P,0.75P\}$. Note that this amounts a total of $5\times 6\times 3 = 90$ configurations. The window size is set to $W=W_{min} = 2000$ for a fair comparison with the baseline case. The number of measurements in dynamics mode is set to $q = \left\lceil S_{ave}\log N\right\rceil = 29$. The number of extra measurements for error control is set to $q^{\prime} = 5 \approx 0.15q$. The average distance and bandwidth saved under each configuration are plotted in Figs.\ref{accuracyVsConfig} and Figs.\ref{BWVsConfig} respectively. The $x$-axis here is a configuration index, which essentially maps to a $(\varepsilon/P,T_{err}/P,\nu/P)$-triplet as specified in Table~\ref{indexToTriplet}. From Table~\ref{indexToTriplet}, we see that configurations $71$ to $90$ meet the average distance target of $0.1$ and in fact, configurations $71$ to $80$ do so while saving up to $15\%$ in control bandwidth.
\begin{figure}[hbt]
\centering
\subfigure[]{\includegraphics[scale=0.5]{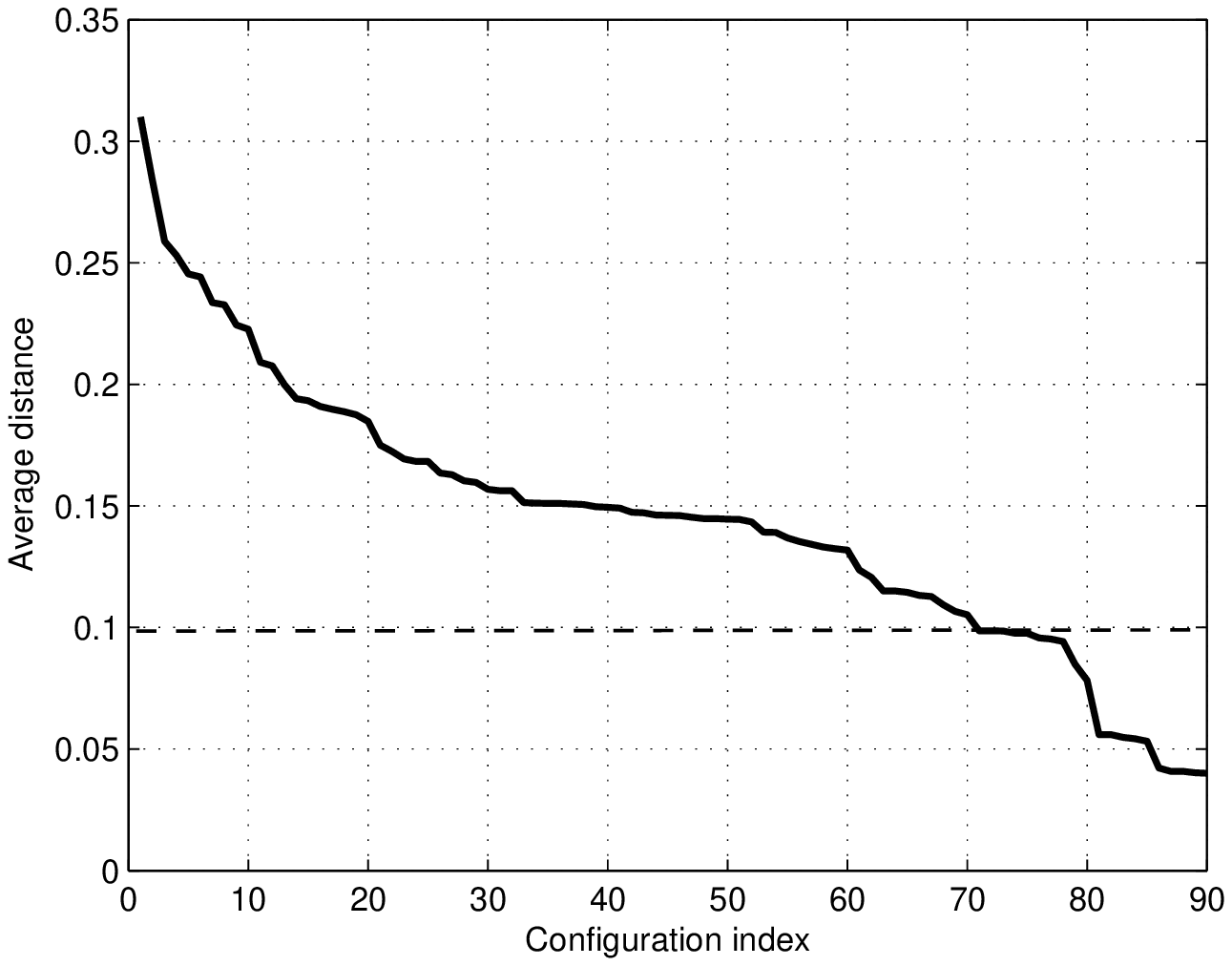}\label{accuracyVsConfig}}
\subfigure[]{\includegraphics[scale=0.5]{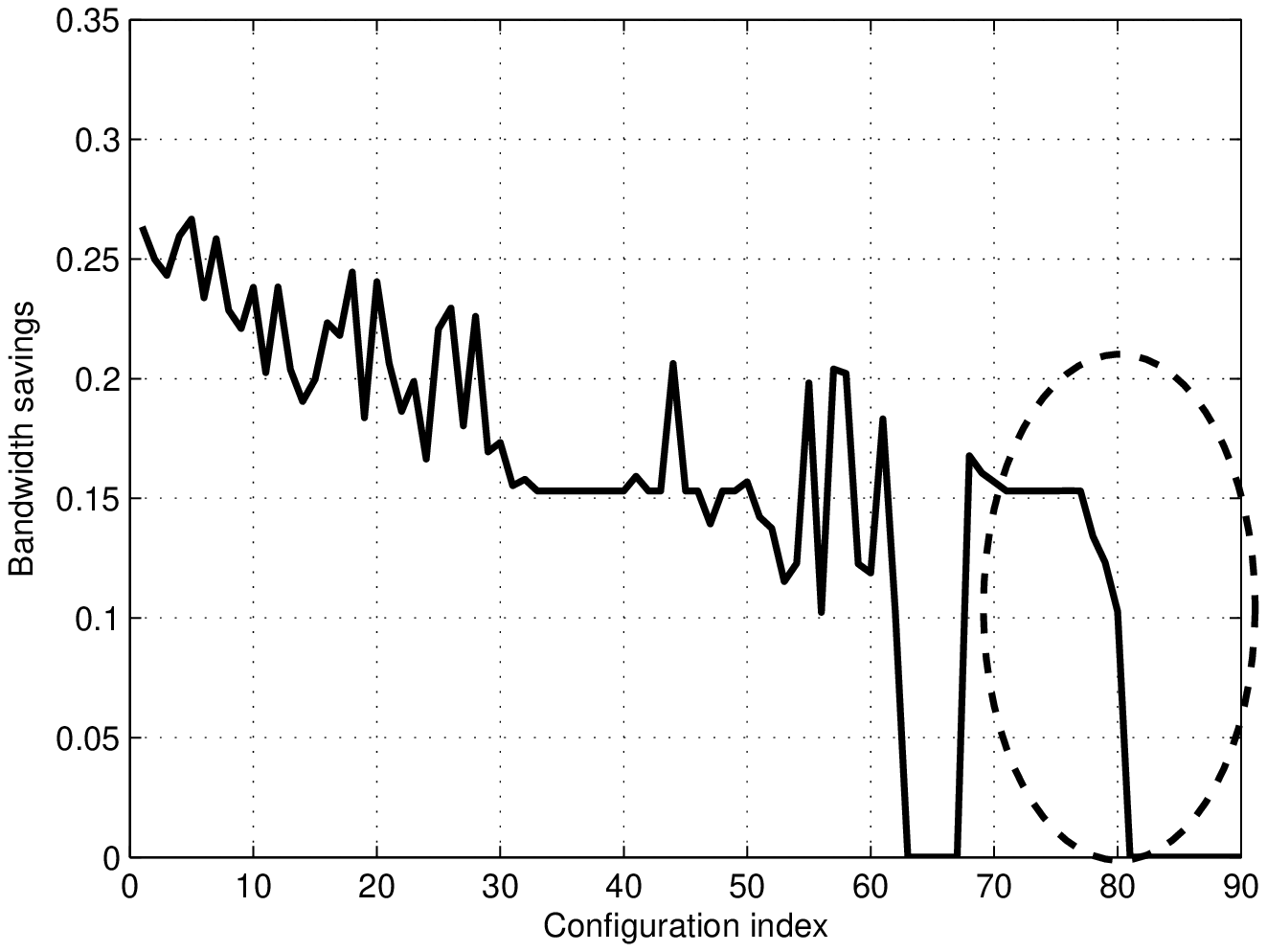}\label{BWVsConfig}}%
\caption{(a) Average distance versus configurations. (b) Bandwidth savings versus configurations.}
\end{figure}

Thus, by exploiting the structure in the emitted signal, we are able to achieve significant savings in control bandwidth over the na\"{i}ve algorithm. Recognizing that the number of parameters that control the algorithm is quite large, we believe that with more extensive tuning/optimization of the various components, e.g., error measurement budget $q^{\prime}$, further savings might be possible, especially as the scale or dimension of the problem grows. Finally, we note that the proposed compressive sampling approach has been applied to (even) more general spatial distribution models such as one where both the sensor and signal nodes are thrown uniformly at random on a square area, in the absence of a grid. The results are as promising but not presented here due to lack of space.

\section{Conclusions}\label{sec: conclusion}

In this paper, we apply ideas from the field of compressive sensing to reduce the control bandwidth in cooperative sensing systems. We are particularly interested in settings where the dynamics of the signals being observed are sparse. This is indeed what one typically expects for systems that exhibit a time-scale separation between the sensing network and the signal emitting process. We prove theoretical RIP-/NSP-based recovery results for path-loss matrices in networks with circular geometries. We then develop a complete cooperative sensing algorithm that addresses key issues such as error propagation. As the proposed algorithm requires many input parameters, we extensively tune these parameters in our numerical experiments. We demonstrate that our approach can provide significant bandwidth savings over the na\"{i}ve linear bandwidth case under more general spatial distributions. Extending the theory to cover these general spatial distributions remains an open problem that deserves attention.

\appendices
\section{Proof of Lemma~\ref{mainlemma1}}\label{mainproof1}
Recall that ${\boldsymbol{\Lambda}}_{{\mathcal C}_{i}}$ represent the sub-matrix of ${\boldsymbol{\Lambda}}_{Q}$ containing the rows specified in ${\mathcal C}_i$. Then, the entries $\{\lambda_{ij}\}$ of sub-matrix ${\boldsymbol{\Lambda}}_{{\mathcal C}_{i}}$ are identically distributed. This follows from the observation that any two sensor nodes on the circle of radius $r_{s,i}$ will perceive the same distribution of signal nodes since the latter nodes are distributed on a circle. Recall that the transformation ${\bf G}={\bf W}{\boldsymbol{\Lambda}}_{Q}$ essentially subtracts rows of ${\boldsymbol{\Lambda}}_{Q}$ corresponding to diametrically opposite users on each circle.

Thus, the columns of ${\bf G}$ are independent and all entries are centered. The next step is to show that the entries $g_{ij}=\lambda_{ij}-\lambda_{(i+1)j}$ are symmetric. This is does not follow immediately from the fact that $\lambda_{ij}$ and $\lambda_{(i+1)j}$ are identically distributed when indices $i$ and $i+1$ come from the same partition since they are not independent. Hence, we appeal to the concept of exchangeable random variables, which is defined below for the special case of a pair of random variables.\\

\noi {\em Definition (Exchangeability)}: Two random variables $X$ and $Y$ are called \textit{exchangeable} if their joint cumulative distribution function (cdf) is symmetric, i.e. if $F_{X,Y}(x,y)=F_{X,Y}(y,x)$.\\

It is known the difference of two identically distributed, exchangeable random variables is indeed symmetric \cite{romano}. Thus, we only need to establish this fact for the case when $\lambda_{ij}$ and $\lambda_{(i+1)j}$ come from the same circle, say ${\mathcal C}_c$, since the definition of ${\bf G}$ in (\ref{Gdef}) precludes any other possibility. To establish that $\lambda_{ij}$ and $\lambda_{(i+1)j}$ are exchangeable, we compute the joint cdf $F_{\lambda_{ij},\lambda_{(i+1)j}}(x,y)$, when $i$ and $i+1$ come from the same circle $c$, as follows
\ben
\ba
F_{\lambda_{ij},\lambda_{(i+1)j}}(x,y)&=&\mbox{\hspace{-0.2cm}}\mbox{Pr}(\lambda_{ij}\leq x,\lambda_{(i+1)j}\leq y)\\
&=&\mbox{\hspace{-0.2cm}}\mbox{Pr}(d_{ij}\geq \frac{1}{x}-K, d_{(i+1)j}\geq \frac{1}{y}-K)\\
&=&\mbox{\hspace{-0.2cm}}\mbox{Pr}\left(d\left(\theta_i,\theta_j\right)\geq \left[\frac{1}{x}-K\right]_+, d\left(\theta_{i+1}\theta_j\right)\geq \right.\\
&&\mbox{\hspace{-0.2cm}}\left.\left[\frac{1}{y}-K\right]_+\right)\mbox{where }[u]_+=\mbox{max}\{u,0\}\\
&=&\mbox{\hspace{-0.2cm}}\frac{1}{2\pi} V({\mathcal R}(\theta_j, \theta_i,x,y)),
\ea
\een
where we have introduced the function
\be
\begin{small}
d^2\left(\theta_m,\theta_n\right)=r_{s,1}^2+r_p^2-2r_p r_{s,1}\mbox{cos }(\theta_m-\theta_n),\bsp i=1,\ldots,\alpha(k),\label{distancedef}
\end{small}
\ee
for notational convenience through the remainder of the proof. The set ${\mathcal R}(\theta_j, \theta_i,x,y)$ is defined as ${\mathcal R}(\theta_j, \theta_i,x,y)=\left\{\theta_j:d\left(\theta_j,\theta_i\right)\geq \left[\frac{1}{x}-K\right]_+, d^2\left(\theta_i,\theta_+\frac{2\pi q}{N_s}\right)\geq \left[\frac{1}{y}-K\right]_+\right\}$ and $V({\mathcal A})$ denotes the volume of the set ${\mathcal A}$. The volume of set ${\mathcal R}(\theta_j, \theta_i,x,y)$ can be expressed as a sum of the volumes corresponding to smaller sets. To that effect, we define $${\mathcal R}_s(\theta,a)=\left\{\theta_j:d\left(\theta_j,\theta\right)\leq \left[\frac{1}{a}-K\right]_+\right\}$$ and can thus write ${\mathcal V}({\mathcal R}(\theta_j,\theta_i,x,y))=2\pi r_p-\left[{\mathcal V}({\mathcal R}_s(\theta_i,x))+{\mathcal V}\left({\mathcal R}_s\left(\theta_i+\frac{2\pi q}{N_s},y\right)\right)-{\mathcal V}\left({\mathcal R}_s\left(\theta_i,x\right) \right.\right.$ $\left.\left.\cap\bsp {\mathcal R}_s\left(\theta_i+\frac{2\pi q}{N_s},y\right) \right)\right].$

From this characterization, we can immediately conclude that ${\mathcal V}({\mathcal R}(r_p,\theta_j, r_{s,c},\theta_i,x,y))={\mathcal V}({\mathcal R}(\theta_j, \theta_i,y,x))$, which gives our the desired result that $\lambda_{ij}$ and $\lambda_{(i+1)j}$ are exchangeable implying that the entries $g_{ij}=\lambda_{ij}-\lambda_{(i+1)j}$ are symmetric for all $(i,j)$. It is clear from the above arguments that exchangeabilty essentially follows due to the uniform distribution of the signal emitters.

The symmetry of $g_{ij}$ is crucial for our next step where we argue that columns of ${\bf A}_{Q}={\bf B}{\bf G}$ remain independent. By definition, $a_{ij}=\frac{1}{\sqrt{\mbox{Var}\{g_{i1}\}}}\beta_{i} g_{ij}$. Thus, to prove that the columns of ${\bf A}_{Q} = \left[{\bf a}_1\bsp {\bf a}_2\ldots {\bf a}_{N_p}\right]$ are independent, we need to show that $a_{ij} \perp {\bf a}_{k}$ for any arbitrary $i,j$ and $k\neq j$. Since the Bernoulli random variables are independent across rows and since $g_{ij} \perp g_{mk},\bsp k\neq j,\bsp m\neq i$, we clearly have that $a_{ij} \perp a_{mk},\bsp k\neq j,\bsp m\neq i$. Thus, we only need to establish that $a_{ij} \perp a_{ik}$ for $k\neq j$ or equivalently that $\beta_i g_{ij} \perp \beta_i g_{ik}$. But this follows from the symmetry of $g_{ij}$ which means that knowledge of $\beta_{i}g_{ij}$ reveals no information about the random variable $\beta_{i}$. Thus, the columns of ${\bf A}_{Q}$ are independent. In addition, there are identically distributed and hence we can now focus on studying the properties of column ${\bf a}_1$ without loss of generality.

The random vector ${\bf a}_1$ is isotropic since ${\mathbb E}[a_{i1}^2]=\frac{1}{{{\mathbb E}[g_{i1}^2]}}{\mathbb E}\left[g_{i1}^2\right] = 1$ and $${\mathbb E}[a_{i1}a_{k1}]=\frac{1}{\sqrt{{{\mathbb E}[g_{i1}^2]}}}\frac{1}{\sqrt{{{\mathbb E}[g_{k1}^2]}}}{\mathbb E}\left[\beta_{i} \beta_{k}\right]{\mathbb E}\left[g_{i1} g_{k1}\right] = 0$$ for $i\neq k$, implying that ${\mathbb E}[|{\bf a}_1^T{\bf x}|^2]=\sum_{i}{\mathbb E}[a_{i1}^2]x_{i}^2+\sum_{i\neq k}{\mathbb E}[a_{i1}a_{k1}]x_ix_k=||{\bf x}||^2.$

To prove that ${\bf a}_1$ is sub-gaussian, we first condition on the position of the first signal emitter $(r_p,\theta_1)$. This will allow us to apply Lemma~\ref{lemmasubgaussian} and Lemma~\ref{lemmasymrvs} since $g_{i1}$ is now completely known thereby making ${\bf a}_{i1}$ a collection of independent random variables. The elements $a_{i1}$ are symmetric and bounded with $|a_{i1}|\leq \frac{|g_{i1}|}{\sqrt{{\mathbb E}[g_{i1}^2]}}$ when conditioned on position $(r_p,\theta_1)$.
\begin{lemma}\label{giboundedbelow}
There exists $M_p>0,\bsp p=0,2,\ldots,q-1$ such that
\be
{\mathbb E}[g_{i1}^2]=M_p, \mbox{ for }p\frac{N_s}{q}< i \leq (p+1)\frac{N_s}{q},\bsp \forall k.
\ee
\end{lemma}
\noi {\em Proof}: By definition, the distribution of $g_{i1}$ for  $p\frac{N_s}{q}< i \leq (p+1)\frac{N_s}{q}$ depends on the distance between the corresponding diametrically opposite sensors on circle $(p+1)$ along with the distribution of the first interfering user. Since this distance always remains the same independent of $i$ and $k$, the result follows.$\hfill\Box$\\

Hence, by Lemma~\ref{lemmasymrvs} and since $|g_{i1}|\leq 1$, $a_{i1}$ is sub-gaussian with $||a_{i1}||_{\psi_2|(r_p,\theta_1)}\leq \frac{1}{M_*^2}$ where $M_*=\mbox{min}_{p=1,\ldots,q} M_p$. Here, we have introduced notation $||\cdot||_{\psi_2|(r_p,\theta_1)}$ to indicate explicitly that we have conditioned on the location of the first interfering user. Now, from Lemma~\ref{lemmasubgaussian}, we conclude that ${\bf a}_{1}$ is a sub-gaussian vector when conditioned on the location of the first signal node with $||{\bf a}_{1}||_{\psi_2|(r_p,\theta_1)}\leq 2C,\bsp C>0$. However, since the sub-gaussian norm $||{\bf a}_{1}||_{\psi_2|(r_p,\theta_1)}$ computed above is independent of $(r_p,\theta_1)$, this implies that ${\bf a}_{1}$ is a sub-gaussian vector with $||{\bf a}_{1}||_{\psi_2}\leq 2C$. This can be seen by applying the Law of Total Probability to the definition of sub-gaussianity in Lemma~\ref{lemmasubgaussian}.$\hfill\Box$

\section{Proof of Lemma~\ref{mainlemma2}}\label{mainproof2}

To show almost-sure convergence of the norm of ${\bf a}_{1}$, we need to prove that $\exists c^*>0$ such that $\mbox{Pr}\left(\mbox{lim}_{k\rightarrow \infty}\left|\frac{1}{k}\sum_{i=1}^{k} a_{i1}^2-c^*\right|=0\right)=\mbox{Pr}\left(\mbox{lim}_{k\rightarrow \infty}\left|\frac{1}{k}\sum_{i=1}^{k} g_{i1}^2-c^*\right|=0\right)=1$ where the probability is computed over the random location $(r_p,\theta),\bsp \theta\sim U[0,2\pi]$. We instead prove the following more general statement that $\exists c^*>0$ such that
\be
\mbox{lim}_{k\rightarrow \infty}\left|\frac{1}{k}\sum_{i=1}^{k} \frac{g_{i1}^2}{{\mathbb E}[g_{i1}^2]}-c^*\right|=0\mbox{ for all } (r_p,\theta),\bsp \theta\in[0,2\pi].\label{ASconvergencelips}
\ee
Note that (\ref{ASconvergencelips}) is a completely deterministic convergence statement in contrast to the earlier probabilistic statement. From the proposed query protocol in Algorithm \ref{alg: feedback}, we see that the number of sensors selected for feedback is a monotonically increasing function of $k$ for all circles. This mean that we can study the convergence of the norm for any one circle (a sub-vector of ${\bf a}_1$) and draw conclusions about the norm concentration of the entire vector ${\bf a}_1$. Consider the first partition of sensors on circle ${\mathcal C}_1$ and let the size of this partition be $\alpha(k)$. Then, we see that $\alpha(k)\rightarrow \infty$ as $k\rightarrow \infty$ and hence, we can shift our focus to proving the property $\mbox{lim}_{k\rightarrow \infty}\left|\frac{1}{\alpha(k)}\sum_{i=1}^{\alpha(k)} \frac{g_{i1}^2}{{\mathbb E}[g_{i1}^2]}-c^*\right|=0,\bsp \forall (r_p,\theta_1),\bsp \theta_1\in[0,2\pi]$. In fact, we show that $c^* = 1$, i.e,
\be
\mbox{lim}_{k\rightarrow \infty}\left|\frac{1}{\alpha(k)}\sum_{i=1}^{\alpha(k)} \frac{g_{i1}^2}{{\mathbb E}[g_{i1}^2]}-1\right|=0,\bsp \forall (r_p,\theta_1),\bsp \theta_1\in[0,2\pi]\label{normConv}
\ee
Recall that the squared-distance between a signal generator located at $(r_p,\theta_1)$ and sensor  node $i$ on ${\mathcal C}_1$ is given by $d^2\left(\theta_i,\theta_1\right)$ where $d(\cdot)$ is defined in (\ref{distancedef})and where $\theta_i=\frac{2\pi i q}{k}$. Now let $f(x,y)=\left(\frac{1}{K+x}-\frac{1}{K+y}\right)^2,\bsp x,y>0$ where $K$ is given in the definition (\ref{PLdefinition1}). Then, $g_{i1}^2$ can equivalently be expressed as $g_{i1}^2 = f\left(d^2\left(\theta_i,\theta_1\right),d^2\left(\theta_i+\frac{\pi}{2},\theta_1\right)\right)$. Now, we proceed to prove (\ref{normConv}) by observing that
\be
\begin{array}{rl}
&\mbox{lim}_{k\rightarrow \infty}\frac{1}{\alpha(k)}\sum_{i=1}^{\alpha(k)} g_{i1}^2\\
=&\mbox{lim}_{k\rightarrow \infty}\frac{1}{\alpha(k)}\sum_{i=1}^{\alpha(k)} f\left(d^2\left(\theta_i,\theta_1\right),d^2\left(\theta_i+\frac{2\pi q}{k},\theta_1\right)\right)\\
=&\frac{1}{2\pi}\int_0^{2\pi} f\left(d^2\left(\theta_i,\theta_1\right),d^2\left(\theta_i+\frac{\pi}{2},\theta_1\right)\right)d\theta_i.\label{num}
\ea
\ee
This claim follows from that fact that the expression on the left is essentially the Riemann sum of the integral on the right. This means that for any given $\varepsilon>0$, we can find $k_{\varepsilon}$ such that when $k\geq k_{\varepsilon}$, we have that $\left|\frac{1}{\alpha(k_{\varepsilon})}\sum_{i=1}^{\alpha(k_{\varepsilon})} g_{i1}^2-\frac{1}{2\pi}\int_0^{2\pi} f\left(d^2\left(\theta_i,\theta\right),d^2\left(\theta_i+\frac{\pi}{2},\theta\right)\right)d\theta_i\right|< \varepsilon$. Next, we calculate
\be
{\mathbb E}[g_{i1}^2]=\frac{1}{2\pi}\int_0^{2\pi} f\left(d^2\left(\theta_i,\theta\right),d^2\left(\theta_i+\frac{\pi}{2},\theta\right)\right)d\theta.\label{denom}
\ee
By substituting (\ref{distancedef}) in (\ref{num}) and (\ref{denom}), we see that the variable of integration may be switched and hence
\be
\ba
&=&\frac{1}{2\pi}\int_0^{2\pi} f\left(d^2\left(\theta_i,\theta\right),d^2\left(\theta_i+\frac{\pi}{2},\theta\right)\right)d\theta\\
&=&\frac{1}{2\pi}\int_0^{2\pi} f\left(d^2\left(\theta_i,\theta\right),d^2\left(\theta_i+\frac{\pi}{2},\theta\right)\right)d\theta_i,\label{equalexpectation}
\ea
\ee
which implies that
\be
\begin{array}{rl}
&\left|\frac{1}{\alpha(k_{\varepsilon})}\sum_{i=1}^{\alpha(k_{\varepsilon})} g_{i1}^2-\frac{1}{2\pi}\int_0^{2\pi} f\left(d^2\left(\theta_i,\theta\right),d^2\left(\theta_i+\frac{\pi}{2},\theta\right)\right)d\theta_i\right| \\
= &\left|\frac{1}{\alpha(k_{\varepsilon})}\sum_{i=1}^{\alpha(k_{\varepsilon})} g_{i1}^2-{\mathbb E}[g_{i1}^2]\right|\\
<&\varepsilon.
\ea
\ee
We can then establish convergence through
\ben
\ba
\left|\frac{1}{\alpha(k)}\sum_{i=1}^{\alpha(k)} \frac{g_{i1}^2}{{\mathbb E}[g_{i1}^2]}-1\right|&=& \frac{1}{{\mathbb E}[g_{i1}^2]}\left|\frac{1}{\alpha(k)}\sum_{i=1}^{\alpha(k)} g_{i1}^2-{\mathbb E}[g_{i1}^2]\right|\\
&<&\frac{1}{{\mathbb E}[g_{i1}^2]}\varepsilon \mbox{ when $k\geq k_{\varepsilon}$.}\\
\ea
\een
The result follows since ${\mathbb E}[g_{i1}^2]$ is bounded below by Lemma~\ref{giboundedbelow}.$\hfill\Box$\\

\renewcommand{\baselinestretch}{0.9}
\begin{center}
\tablefirsthead{%
\hline
\multicolumn{1}{|c|}{Config.} &
\multicolumn{1}{c|}{$(\varepsilon/P,T_{err}/P,\nu/P)$-triplet} &
\multicolumn{1}{c|}{Distance} &
\multicolumn{1}{c|}{Bandwidth} \\
\multicolumn{1}{|c|}{index} &
\multicolumn{1}{c|}{} &
\multicolumn{1}{c|}{} &
\multicolumn{1}{c|}{savings} \\
\hline}
\tablehead{%
\hline
\multicolumn{4}{|l|}{\small\sl continued from previous column}\\
\hline
\multicolumn{1}{|c|}{Config.} &
\multicolumn{1}{c|}{$(\varepsilon/P,T_{err}/P,\nu/P)$-triplet} &
\multicolumn{1}{c|}{Distance} &
\multicolumn{1}{c|}{Bandwidth} \\
\multicolumn{1}{|c|}{index} &
\multicolumn{1}{c|}{} &
\multicolumn{1}{c|}{} &
\multicolumn{1}{c|}{savings} \\
\hline}
\tabletail{%
\hline
\multicolumn{4}{|r|}{\small\sl continued on next column}\\
\hline}
\tablelasttail{\hline}
\bottomcaption{Mapping from configuration index to parameter triplet}\label{indexToTriplet}
\begin{footnotesize}
\begin{supertabular}{|c|c|c|c|}
\hline
1&(3,0.25,0.8)&0.31&0.2634\\\hline
2&(2,0.25,0.8)&0.2838&0.2499\\\hline
3&(3,0.25,0.5)&0.2588&0.2432\\\hline
4&(3,0.75,0.8)&0.253&0.2597\\\hline
5&(3,0.5,0.8)&0.2454&0.2667\\\hline
6&(2,0.25,0.5)&0.2442&0.2338\\\hline
7&(2,0.5,0.8)&0.2336&0.2584\\\hline
8&(3,0.25,0.4)&0.2327&0.2285\\\hline
9&(2,0.25,0.4)&0.2244&0.221\\\hline
10&(2,0.75,0.8)&0.2227&0.2381\\\hline
11&(1,0.25,0.8)&0.2091&0.2025\\\hline
12&(3,0.75,0.5)&0.2076&0.2383\\\hline
13&(3,0.25,0.3)&0.1999&0.2037\\\hline
14&(1,0.25,0.5)&0.1941&0.1904\\\hline
15&(2,0.25,0.3)&0.1933&0.1997\\\hline
16&(3,0.75,0.4)&0.1909&0.2233\\\hline
17&(2,0.75,0.5)&0.1898&0.218\\\hline
18&(3,0.5,0.5)&0.1888&0.2445\\\hline
19&(1,0.25,0.4)&0.1875&0.1836\\\hline
20&(2,0.5,0.5)&0.1848&0.2405\\\hline
21&(2,0.75,0.4)&0.175&0.2064\\\hline
22&(1,0.75,0.8)&0.1724&0.1863\\\hline
23&(3,0.75,0.3)&0.1693&0.1988\\\hline
24&(1,0.25,0.3)&0.1683&0.1663\\\hline
25&(1,0.5,0.8)&0.1683&0.2207\\\hline
26&(3,0.5,0.4)&0.1635&0.2295\\\hline
27&(2,0.75,0.3)&0.1628&0.1802\\\hline
28&(2,0.5,0.4)&0.1603&0.2261\\\hline
29&(1,0.75,0.5)&0.1597&0.1694\\\hline
30&(1,0.75,0.4)&0.1568&0.1733\\\hline
31&(0.5,0.25,0.4)&0.1562&0.1552\\\hline
32&(0.5,0.25,0.8)&0.1562&0.1579\\\hline
33&(0.1,0.25,0.4)&0.1514&0.1531\\\hline
34&(1,0.25,0.2)&0.1511&0.1531\\\hline
35&(0.1,0.25,0.2)&0.151&0.1531\\\hline
36&(3,0.25,0.2)&0.151&0.1531\\\hline
37&(0.1,0.75,0.8)&0.1508&0.1531\\\hline
38&(0.1,0.25,0.8)&0.1506&0.1531\\\hline
39&(2,0.25,0.2)&0.1497&0.1531\\\hline
40&(0.5,0.25,0.2)&0.1494&0.1531\\\hline
41&(0.5,0.75,0.8)&0.1491&0.1592\\\hline
42&(0.1,0.75,0.4)&0.1474&0.1531\\\hline
43&(0.5,0.75,0.2)&0.1472&0.1531\\\hline
44&(1,0.5,0.5)&0.1463&0.2063\\\hline
45&(2,0.75,0.2)&0.1461&0.1531\\\hline
46&(1,0.75,0.2)&0.146&0.1531\\\hline
47&(0.5,0.25,0.5)&0.1453&0.1393\\\hline
48&(0.1,0.75,0.2)&0.1448&0.1531\\\hline
49&(3,0.75,0.2)&0.1448&0.1531\\\hline
50&(0.5,0.75,0.4)&0.1446&0.1569\\\hline
51&(1,0.75,0.3)&0.1444&0.1422\\\hline
52&(0.5,0.75,0.5)&0.1434&0.1374\\\hline
53&(0.5,0.75,0.3)&0.1392&0.1151\\\hline
54&(0.1,0.75,0.5)&0.1391&0.1228\\\hline
55&(1,0.5,0.4)&0.1368&0.1982\\\hline
56&(0.1,0.75,0.3)&0.1354&0.1024\\\hline
57&(3,0.5,0.3)&0.1342&0.204\\\hline
58&(2,0.5,0.3)&0.1331&0.2021\\\hline
59&(0.1,0.25,0.5)&0.1324&0.1225\\\hline
60&(0.5,0.25,0.3)&0.1318&0.1187\\\hline
61&(1,0.5,0.3)&0.1236&0.1832\\\hline
62&(0.1,0.25,0.3)&0.1206&0.1021\\\hline
63&(0.1,0.75,0.1)&0.115&0\\\hline
64&(3,0.75,0.1)&0.115&0\\\hline
65&(1,0.75,0.1)&0.1145&0\\\hline
66&(2,0.75,0.1)&0.1132&0\\\hline
67&(0.5,0.75,0.1)&0.1127&0\\\hline
68&(0.5,0.5,0.8)&0.1093&0.1677\\\hline
69&(0.5,0.5,0.4)&0.1066&0.1606\\\hline
70&(0.5,0.5,0.5)&0.1053&0.1569\\\hline
71&(1,0.5,0.2)&0.0987&0.1531\\\hline
72&(2,0.5,0.2)&0.0987&0.1531\\\hline
73&(0.5,0.5,0.2)&0.0986&0.1531\\\hline
74&(0.1,0.5,0.2)&0.0976&0.1531\\\hline
75&(3,0.5,0.2)&0.0976&0.1531\\\hline
76&(0.1,0.5,0.8)&0.0957&0.1531\\\hline
77&(0.1,0.5,0.4)&0.0952&0.1531\\\hline
78&(0.5,0.5,0.3)&0.0942&0.1341\\\hline
79&(0.1,0.5,0.5)&0.0849&0.1232\\\hline
80&(0.1,0.5,0.3)&0.0782&0.1027\\\hline
81&(0.1,0.25,0.1)&0.056&0\\\hline
82&(3,0.25,0.1)&0.056&0\\\hline
83&(2,0.25,0.1)&0.0548&0\\\hline
84&(0.5,0.25,0.1)&0.0543&0\\\hline
85&(1,0.25,0.1)&0.0532&0\\\hline
86&(2,0.5,0.1)&0.0422&0\\\hline
87&(0.1,0.5,0.1)&0.0408&0\\\hline
88&(3,0.5,0.1)&0.0408&0\\\hline
89&(0.5,0.5,0.1)&0.0403&0\\\hline
90&(1,0.5,0.1)&0.0401&0\\\hline
\end{supertabular}
\end{footnotesize}
\end{center}

\end{document}